\documentclass{iau}
\usepackage{graphicx}

\title[Pulsations in evolved hot stars] %% give here short title %%
{Pulsations as a mass-loss trigger \\ in evolved hot stars\thanks{Based on 
observations acquired at the Ond\v{r}ejov Observatory, Czech 
Republic, the Dominion Astrophysical Observatory, Herzberg Institute of 
Astrophysics, National Research Council of Canada, and with the HERMES spectrograph, 
which is supported by the Fund for Scientific Research of Flanders (FWO), 
Belgium, the Research Council of K.U.Leuven, Belgium, the Fonds National 
Recherches Scientific (FNRS), Belgium, the Royal Observatory of Belgium, the 
Observatoire de Gen\`{e}ve, Switzerland and the Th\"{u}ringer Landessternwarte 
Tautenburg, Germany.}}

\author[Michaela Kraus et al.]   %% give here short author list %%
{Michaela Kraus$^1$, Dieter H. Nickeler$^1$, Maximiliano Haucke$^2$, \\ 
Lydia Cidale$^{2,3}$, Roberto Venero$^{2,3}$, Marcelo Borges Fernandes$^4$, \\
Sanja Tomi\'{c}$^5$ \and Michel Cur\'{e}$^6$}

\affiliation{$^1$Astronomick\'y \'ustav, Akademie v\v{e}d \v{C}esk\'e republiky, \\ Fri\v{c}ova 298, 251\,65 Ond\v{r}ejov, Czech Republic\\ email: {\tt kraus@sunstel.asu.cas.cz} \\[\affilskip]
$^2$Departamento de Espectroscop\'ia Estelar, Facultad de Ciencias
Astron\'omicas y Geof\'isicas, \\ Universidad Nacional de La Plata, Paseo del
Bosque s/n, B1900FWA, La Plata, Argentina \\[\affilskip]
$^3$Instituto de Astrof\'isica de La Plata, CCT La Plata, CONICET-UNLP, \\ Paseo del Bosque s/n, B1900FWA, La Plata, Argentina \\[\affilskip]
$^4$Observat\'orio Nacional, \\ Rua General Jos\'e Cristino 77, 20921-400 S\~ao Cristov\~ao, Rio de Janeiro, Brazil \\[\affilskip]
$^5$Department of Astronomy, Faculty of Mathematics, University of Belgrade, \\ Studentski trg 16, 11000 Belgrade, Serbia  \\[\affilskip]
$^6$Departmento de F\'isica y Astronom\'ia, Facultad de Ciencias, Universidad de Valpara\'iso, \\
Av. Gran Breta\~na 1111, Casilla 5030, Valpara\'iso, Chile 
}

\pubyear{2014}
\volume{301}  %% insert here IAU Symposium No.
\pagerange{--}
\jname{Precision Asteroseismology}
\editors{W. Chaplin, J. Guzik, G. Handler \& A. Pigulski, eds.}
\begin{document}

\maketitle

\begin{abstract}
During the post-main sequence evolution massive stars pass through several
short-lived phases, in which they experience enhanced mass loss in the form of 
clumped winds and mass ejection events of unclear origin. The discovery that 
stars populating the blue, luminous part of the Hertzsprung-Russell diagram 
can pulsate hence 
suggests that stellar pulsations might influence or trigger enhanced mass loss 
and eruptions. We present recent results for two objects in different phases:
a B[e] star at the end of the main sequence and a B-type supergiant.
\end{abstract}

\firstsection % if your document starts with a section,
\section{Introduction}

The post-main sequence evolution of massive stars is one of the major unsolved 
problems in massive star research. Massive stars can pass through several 
short-lived phases, in which they lose tremendous amounts of mass via enhanced 
mass-loss and eruptive mass ejection events of yet unknown origin. 
During the classical Blue Supergiant (BSG) stage mass-loss occurs
via line-driven winds. The mass-loss rates involved are still uncertain and 
strongly depend on whether the winds emerge smoothly from the stellar surface 
or whether instabilities occur at the base of the wind, which result in the 
formation of clumpy structures. The cause of such instabilities is still 
unclear, but stellar pulsations, which were recently found in a couple of BSGs 
(e.g., \cite[Saio et al. 2006]{Saio_etal06}, \cite[Lefever et al. 
2007]{Lefever_etal07}, \cite[Kraus et al. 2012]{Kraus_etal12}) might 
influence, and maybe even trigger, enhanced mass loss from evolved hot stars.   
Here we present recent results for a B[e] star and a BSG.

\section{The Galactic B[e] star HD\,50138}

HD\,50138, as a member of the B[e] stars, is surrounded by high-density 
material giving rise to strong Balmer and forbidden emission lines, and a 
circumstellar dusty disk ($i = 56\pm 4^{\circ}$) resolved by interferometry 
(\cite[Borges Fernandes et al. 2011]{BorgesFernandes_etal11}). It experienced 
outbursts and shell-phases in the past, and its location at the end of (or 
slightly beyond) the main-sequence evolution (\cite[Borges Fernandes et al. 
2009]{BorgesFernandes_etal09}), close to confirmed pulsating Be stars, 
suggests pulsations as possible trigger for the outbursts.

\begin{figure}[t]
\begin{center}
 \includegraphics[width=12cm]{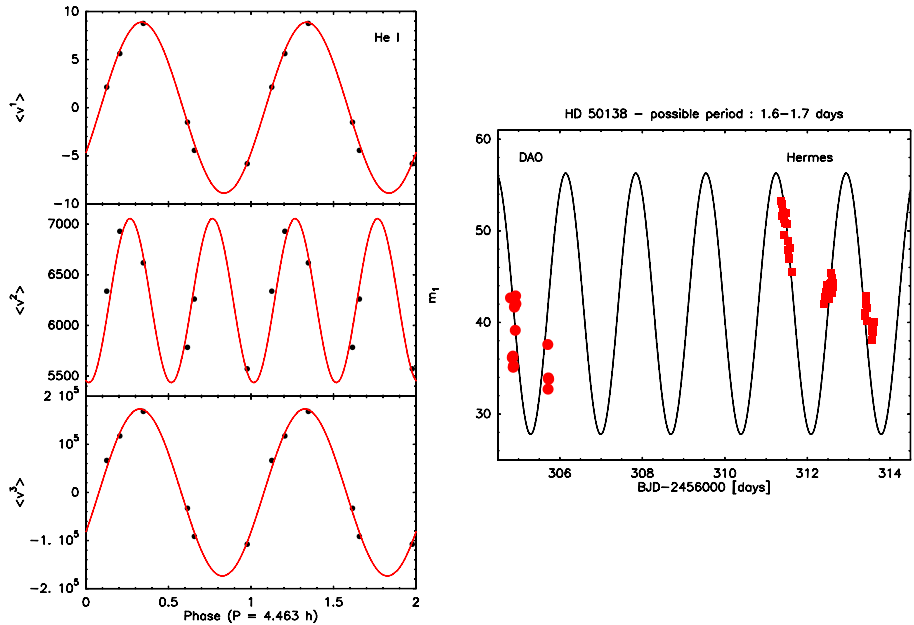}
 \caption{First three moments of the He{\sc i} 4026\,\AA~line phased to the 
possible short-term period of 4.463\,h (left). Radial velocity (first moment) 
variations in the combined data sets from DAO and HERMES imply the existence of 
an additional longer period with larger amplitude (right).}
   \label{fig1}
\end{center}
\end{figure}

First indications for pulsational activity in the atmosphere of HD\,50138 were
found from a sample of high-resolution spectra obtained during different 
observing runs at the 1.2-m Mercator (HERMES) and DAO telescopes. The data
showed strong night-to-night variability in all photospheric and wind lines 
(\cite[Borges Fernandes et al. 2012]{BorgesFernandes_etal12}). In addition, the 
photospheric lines displayed a large broadening component of 30 -
40\,km\,s$^{-1}$ in excess to the stellar rotational broadening ($v \sin i = 
74.7\pm 0.8$\,km\,s$^{-1}$). Such high values of excess broadening 
(referred to as `macroturbulence') is well known from BSGs and is attributed to 
pulsational activity (\cite[Aerts et al. 2009]{Aerts_etal09}). Application of 
the moment method (\cite[Aerts et al. 1992]{Aerts_etal92}; \cite[North \& 
Paltani 1994]{NorthPaltani94}) to the photospheric He\,{\sc i} and Si\,{\sc ii} 
lines suggests the presence of two possible periods (4.463\,h and 1.6-1.7\,d). 
The results for the He\,{\sc i} 4026\,\AA~line are depicted in 
Fig.\,\ref{fig1}. If these periods are confirmed, HD\,50138 will be the first 
pulsating B[e] star, and as such it will provide a very important milestone for 
our understanding of the triggering mechanism leading to mass ejection events 
in B[e] stars.

\section{The blue supergiant star 55\,Cyg = HD\,198478}

A second group of evolved hot and massive stars discussed during this meeting
(see Godart et al., this volume) are BSGs. Members of this class were long
known to display strong photometric and spectroscopic variability, and the
profiles of their photospheric lines contain large contributions from 
macroturbulent broadening in excess to rotational broadening (e.g., 
\cite[Markova \& Puls 2008]{MarkovaPuls08}), indicating stellar pulsational 
activity. In fact, recent theoretical investigations by \cite[Saio et al. 
(2006)]{Saio_etal06} revealed the presence of a new instability domain in the 
HRD covering the location of the BSGs.

\begin{figure}[t]
\begin{center}
 \includegraphics[width=11cm]{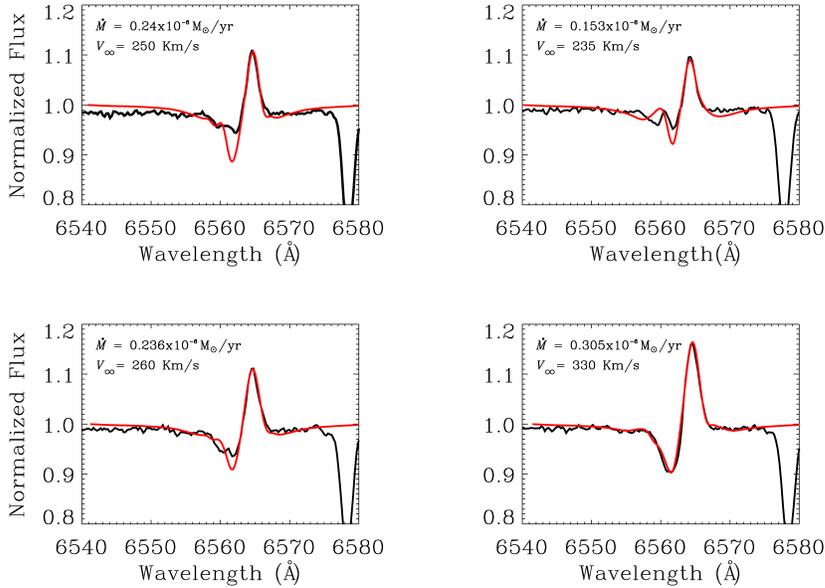}
 \caption{Fits to the H$\alpha$ profile observed within four consecutive 
nights (Sep 21-24, 2010). Note the strong increase in $\dot{M}$ and 
$v_{\infty}$ from the 2nd to the last night.}
   \label{fig2}
\end{center}
\end{figure}

We study a sample of bright northern BSGs
located within this instability domain using the Perek 2-m telescope at 
Ond\v{r}ejov Observatory. One of the objects we are surveying is the early 
B-type supergiant 55\,Cyg (HD\,198478). While its stellar parameters have been 
determined accurately from optical spectroscopy ($T_{\rm eff} = 17\,500 \pm 
500$\,K; $\log L/L_{\odot} = 5.1\pm 0.2$; $v \sin i = 37\pm 2$\,km\,s$^{-1}$; 
$v_{\rm macro}=53$\,km\,s$^{-1}$; \cite[Markova \& Puls 2008]{MarkovaPuls08}), 
the situation is less clear regarding the wind parameters. Mass-loss rates and 
terminal wind velocities are typically obtained from the emission component of 
the H$\alpha$ line. However, the H$\alpha$ line displays strong 
night-to-night variability. From our long-term observations we found a zoo of 
profile shapes ranging from P\,Cygni over pure single emission,
almost complete disappearance, to double- or multiple-peaked, and no
cyclic variation was found over a total of 25 consecutive observing nights.
Consequently, modeling the emission component of observations taken in
different nights delivered different sets of wind parameters. 

So far we collected a total of 339 spectra in the H$\alpha$ region distributed 
over 59 nights between August 2009 and August 2013. The spectral coverage is 
6270-6730\,\AA~with a resolution of $R\simeq 13\,000$. Of these, we modeled the 
H$\alpha$ profile from 32 different nights using the NLTE code FASTWIND 
(\cite[Puls et al. 2005]{Puls_etal05}) to obtain the wind parameters. We found 
that the mass-loss rate, $\dot{M}$, and terminal wind velocity, $v_{\infty}$, 
change simultaneously with large night-to-night variability (Fig.\,\ref{fig2}). 
The value in both parameters spreads over more than a factor of three: 
$\dot{M} = (1.4 - 4.3)\cdot 10^{-7}$\,M$_{\odot}$/yr and $v_{\infty} = 180 - 
700$\,km/s. 

\begin{figure}[t]
\begin{center}
 \includegraphics[width=11.5cm]{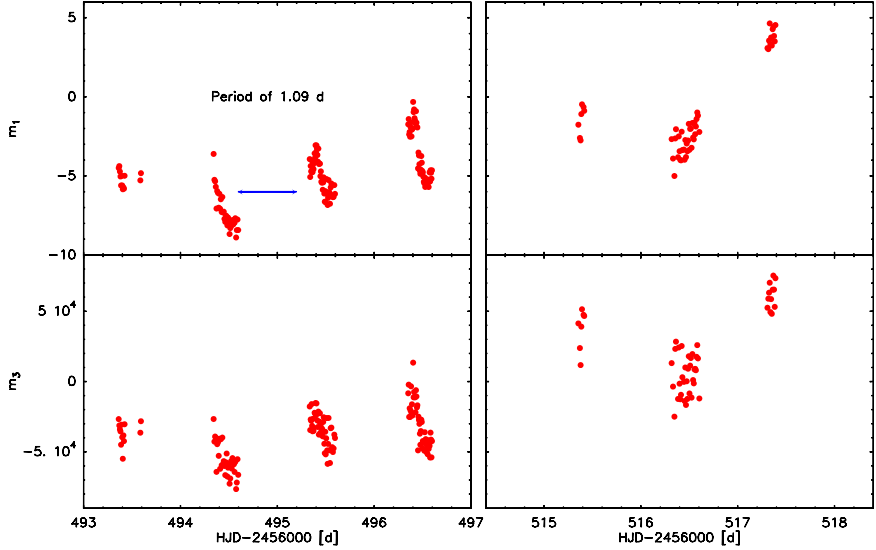}
 \caption{First (top) and third (bottom) moments 
of the He{\sc i} 6678\,\AA~line showing identical variation typical for 
pulsations. The moments were computed from time-series within four (left) and 
three (right) consecutive nights. Both observing epochs suggest
a possible period of 1.09\,d. The shift in radial velocity
between the two epochs indicates additional superimposed period(s).}
   \label{fig3}
\end{center}
\end{figure}

From the moment analysis of time-series within four and three consecutive 
nights of the He{\sc i} 6678\,\AA~line we obtained a possible pulsation period
of 1.09\,d (Fig.\,\ref{fig3}). The shift in radial velocity between the first 
and second set of time series suggests that the 1.09\,d period is superimposed 
on a second (and probably more) period(s). A proper period and mode analysis 
(work in progress) is necessary to confirm the identifications.

\begin{acknowledgements}
M.K. and D.H.N. acknowledge financial support from GA\v{C}R under
grant number P209/11/1198. Financial support for International Cooperation of
the Czech Republic (M\v{S}MT, 7AMB12AR021) and Argentina (Mincyt-Meys,
ARC/11/10) 
is acknowledged. The Astronomical 
Institute Ond\v{r}ejov is supported by the project RVO:67985815. 
L.C. acknowledges financial support from CONICET (PIP 0300) and the
Agencia (pr\'estamo BID, PICT 2011/0885).
\end{acknowledgements}

\end{document}